\documentclass[twocolumn,showpacs,prc,superscriptaddress,amsmath,amssymb]{revtex4-1}
\usepackage{multirow}
\usepackage{graphicx}% Include figure files
\usepackage{dcolumn}% Align table columns on decimal point
\usepackage{bm}% bold math
\usepackage{soul}
\usepackage{color}
\begin{document}

\preprint{APS/123-QED}

%\title{On the question of two-neutron radioactivity}
\title{Study of two-neutron radioactivity in the decay of $^{26}$O}

\author{Z. Kohley}
 \email{kohley@nscl.msu.edu}
\affiliation{National Superconducting Cyclotron Laboratory, Michigan State University, East Lansing, Michigan 48824, USA}
\affiliation{Department of Chemistry, Michigan State University, East Lansing, Michigan 48824, USA}
\author{T.~Baumann}
	\affiliation{National Superconducting Cyclotron Laboratory, Michigan State University, East Lansing, Michigan 48824, USA}
\author{D.~Bazin}
	\affiliation{National Superconducting Cyclotron Laboratory, Michigan State University, East Lansing, Michigan 48824, USA}
\author{G.~Christian}
	\affiliation{National Superconducting Cyclotron Laboratory, Michigan State University, East Lansing, Michigan 48824, USA}
	\affiliation{Department of Physics \& Astronomy, Michigan State University, East Lansing, Michigan 48824, USA}
\author{P.A.~DeYoung}
	\affiliation{Department of Physics, Hope College, Holland, Michigan 49423, USA}
\author{J.E.~Finck}
    \affiliation{Department of Physics, Central Michigan University, Mt. Pleasant, Michigan, 48859, USA}
\author{N.~Frank}
	\affiliation{Department of Physics \& Astronomy, Augustana College, Rock Island, Illinois, 61201, USA}
\author{M.~Jones}
	\affiliation{National Superconducting Cyclotron Laboratory, Michigan State University, East Lansing, Michigan 48824, USA}
	\affiliation{Department of Physics \& Astronomy, Michigan State University, East Lansing, Michigan 48824, USA}
\author{E.~Lunderberg}
	\affiliation{Department of Physics, Hope College, Holland, Michigan 49423, USA}
\author{B.~Luther}
    \affiliation{Department of Physics, Concordia College, Moorhead, Minnesota 56562, USA}
\author{S.~Mosby}
	\affiliation{National Superconducting Cyclotron Laboratory, Michigan State University, East Lansing, Michigan 48824, USA}
	\affiliation{Department of Physics \& Astronomy, Michigan State University, East Lansing, Michigan 48824, USA}
\author{T.~Nagi}
	\affiliation{Department of Physics, Hope College, Holland, Michigan 49423, USA}
\author{J.~K.~Smith}
	\affiliation{National Superconducting Cyclotron Laboratory, Michigan State University, East Lansing, Michigan 48824, USA}
	\affiliation{Department of Physics \& Astronomy, Michigan State University, East Lansing, Michigan 48824, USA}
\author{J.~Snyder}
	\affiliation{National Superconducting Cyclotron Laboratory, Michigan State University, East Lansing, Michigan 48824, USA}
	\affiliation{Department of Physics \& Astronomy, Michigan State University, East Lansing, Michigan 48824, USA}
\author{A.~Spyrou}
	\affiliation{National Superconducting Cyclotron Laboratory, Michigan State University, East Lansing, Michigan 48824, USA}
	\affiliation{Department of Physics \& Astronomy, Michigan State University, East Lansing, Michigan 48824, USA}
\author{M.~Thoennessen}
	\affiliation{National Superconducting Cyclotron Laboratory, Michigan State University, East Lansing, Michigan 48824, USA}
	\affiliation{Department of Physics \& Astronomy, Michigan State University, East Lansing, Michigan 48824, USA}

\date{\today}% It is always \today, today,

\begin{abstract}
A new technique was developed to measure the lifetimes of neutron unbound nuclei in the picosecond range.  The decay of $^{26}$O$\rightarrow$$^{24}\mathrm{O}+n+n$ was examined as it had been predicted to have an appreciable lifetime due to the unique structure of the neutron-rich oxygen isotopes.  The half-life of $^{26}$O was extracted as 4.5$^{+1.1}_{-1.5}~(stat.) \pm 3 (sys.)$~ps.  This corresponds to $^{26}$O having a finite lifetime at an 82$\%$ confidence level and, thus, suggests the possibility of two-neutron radioactivity.

%While the lower limit can not be resolved with the same confidence, it does suggest the strong possibility for a finite lifetime of $^{26}$O and, thus, the observation of two-neutron radioactivity.

  %This measurement provides a stringent upper-limit and suggests the possibility

%, which provides a stringent upper-limit and hints at the possibility of two-neutron radioactivity.

%Thus, a stringent upper-limit on the half-life of 7.25~ps can be assigned with a 95$\%$ confidence level.

%providing the first evidence for two-neutron radioactivity.  7.25 ps 95% confidence level
\end{abstract}


\maketitle
\par
Drastic changes in the structure, properties, and available decay-modes of isotopes with extreme neutron-to-proton ratios have been observed in comparison to their stable counterparts~\cite{BAUMANN12,Pfu12,Tho04,Han01,BROWN01,Ots01}.  In 1960 Goldansky predicted that the unique properties of very proton-rich nuclei would produce scenarios in which one- and two-proton radioactivity could be observed~\cite{Gold60}.  These exotic modes of radioactivity were later verified through measurements of the one-proton decay of $^{151}$Lu~\cite{Hof82} and $^{147}$Tm~\cite{Klep82} and the two-proton decay of $^{45}$Fe~\cite{Pfu02,Gio02}. In particular, a measurement by Miernik \emph{et al.} allowed for the exotic two-proton radioactivity of $^{45}$Fe to be fully characterized for the first time providing insight into the three-body structure of this proton dripline nucleus~\cite{Mie07}.  New modes of radioactivity have also been observed for neutron-rich nuclei including $\beta$-delayed two-~\cite{Asu79}, three-~\cite{Azu80}, and four-neutron decays~\cite{Duf88}.

\par
Recently, Grigorenko \emph{et al.} calculated the lifetimes of one-, two-, and four-neutron decays from unbound nuclei~\cite{Gri11}.  While long lifetimes for proton decay can arise from the presence of both the Coulomb and angular momentum barriers, the neutron decay must only overcome the angular momentum barrier and will therefore have a much shorter lifetime.  However, Grigorenko \emph{et al.} showed that it would be possible for some neutron-rich unbound nuclei to have long lifetimes, reaching the limit of radioactivity, depending on the nuclear structure~\cite{Gri11}.  While an exact limit for a lifetime to be considered radioactivity does not exist, different arguments have been presented suggesting a lower limit between 10$^{-14}$~s and 10$^{-12}$~s (see discussion in Refs.~\cite{Pfu12} and~\cite{Tho04}).  For the two-neutron unbound $^{26}$O a lifetime on the order of 10$^{-12}$~s was predicted for a ground-state resonance energy of about 150~keV and a pure $\nu$($d_{3/2}$)$^{2}$ configuration for the valence neutrons~\cite{Gri11}.

\par
The recent measurement of the $^{26}$O ground state resonance at $E_{decay}$ = 150$^{+50}_{-150}$~keV by Lunderberg \emph{et al.} opened up the exciting possibility for a new type of radioactivity to be discovered~\cite{LUN12}.  This limit was confirmed by the GSI-LAND group which determined the $^{26}$O ground state to be unbound by less than 120~keV~\cite{Cae12}.  Upper-limits have already been placed on the lifetime of $^{26}$O from previous experiments.  Searches for $^{26}$O using fragment separators placed a lifetime limit of $<$~200~ns (roughly the flight time through the separator) based on the non-observation of the bound nucleus~\cite{Gui90,Tar97}.  The GSI-LAND group improved that limit by an order of magnitude to $<$~5.7~ns based on the reconstruction of an $^{26}$O fragment which decayed in flight outside the target~\cite{Cae12}.  The $^{26}$O lifetime is therefore too short to be observed by traditional implantation/decay methods, where lifetimes down to several hundred nanoseconds have been achieved using digital electronics~\cite{Lid06}, and must be measured in-flight.  For two-proton emission, reconstruction of the decay vertex by Mukha \emph{et al.}~\cite{Muk07} and an adaption of the Recoil Distance Doppler Shift (RDDS) technique by Voss \emph{et al.}~\cite{Voss12} were used to measure the lifetime in-flight of $^{19}$Mg.

\par
In this letter, the lifetime of $^{26}$O is extracted using a novel technique based on the velocity difference of the emitted neutrons and residual charged fragment traveling through a thick target.  This approach is a variance on the Doppler-Shift Attenuation Method (DSAM) in which an excited nucleus is slowed in a solid material resulting in a distribution of Doppler shifted $\gamma$-ray energies that can be related to the lifetime~\cite{Sch68,Nol79,Dew12}.  This concept is extended in the present work for nuclei which decay by neutron emission.  The analysis is completed using the experimental data from the work of Lunderberg \emph{et al.}~\cite{LUN12}, in which the $^{26}$O ground state resonance was originally measured.

\par
Since the experimental details have been provided previously in Ref.~\cite{LUN12}, only a brief overview is presented.  The $^{26}$O was produced using a one-proton knockout reaction from a 82 MeV/u $^{27}$F beam produced at the National Superconducting Cyclotron Laboratory at Michigan State University.  To produce the $^{27}$F beam, a 140 MeV/u primary beam of $^{48}$Ca bombarded a 1316 mg/cm$^{2}$ $^{9}$Be production target.  The A1900 fragment separator~\cite{Mor03} was used to select the desired $^{27}$F fragments which were then impinged on a 705 mg/cm$^{2}$ $^{9}$Be reaction target in the experimental vault. The $^{26}$O$\rightarrow^{24}\mathrm{O}+n+n$ decay was measured using the Modular Neutron Array (MoNA) and the 4~Tm superconducting dipole magnet~\cite{SWEEPER}.  The dipole magnet bent the charged fragments about 43$^{\circ}$ into a suite of charged particle detectors, which allowed for the mass, charge, kinetic energy, and angle of the charged particle to be reconstructed from its track through the magnet~\cite{Chr12}.  MoNA was placed 6.05 m from the reaction target and provided the measurement of the velocity and angle of the neutrons.

\par
The three-body decay energy of the $^{24}\mathrm{O}+n+n$ system was calculated as $E_{\mathrm{decay}} = M_{^{26}\mathrm{O}} - M_{^{24}\mathrm{O}} - 2M_{n}$, where $M_{^{26}\mathrm{O}}$ ($M_{^{24}\mathrm{O}}$) is the mass of $^{26}$O ($^{24}$O) and $M_{\mathrm{n}}$ is the neutron mass.  The invariant mass, $M_{^{26}\mathrm{O}}$, was calculated from the experimentally measured four-momenta of the $^{24}$O and two neutrons. The three-body decay spectra requires a triple coincidence of two interactions in MoNA that pass the causality cuts and a $^{24}$O fragment.  The causality cuts are used to select true $2n$ events from multiple scattering of a single neutron and are discussed in detail in Ref.~\cite{LUN12}.

\begin{figure}
\includegraphics[width=0.40\textwidth]{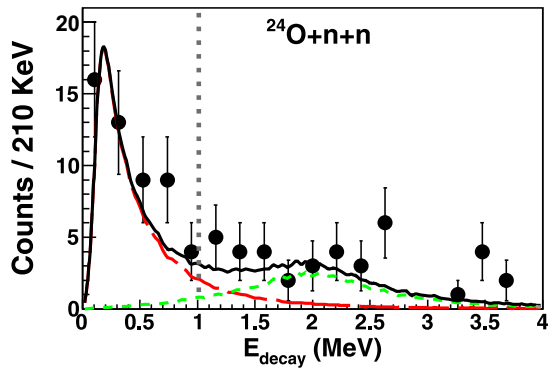}
\caption{\label{f:edecay} (Color online) Experimental $^{24}\mathrm{O}+n+n$ decay energy spectrum (solid black points) with causality cuts applied is compared with the Monte Carlo simulation (solid black line) with two components: (red long-dashed line) the $^{26}$O ground state resonance and (green short-dashed line) the first excited state.  The vertical dotted line represents the selection of $^{26}$O events used in the analysis.}
\end{figure}

\par
A detailed Monte Carlo simulation was used to fit the experimental spectrum as described in Ref.~\cite{LUN12}.  The simulation included all relevant components of the experimental setup.  In particular, special care was taken in reproducing the neutron interaction observables in MoNA using the \texttt{\sc Geant4} framework with the custom neutron interaction model \textsc{menate\_r}~\cite{Koh12}.  As shown in Fig.~\ref{f:edecay}, the experimental three-body decay spectrum was well reproduced by the Monte Carlo simulation including the decay from both the ground state (red long-dashed line) and first excited state (green dashed line).  It is important to note that the ground state resonance was determined from a fit of the data~\cite{LUN12} while the placement of the first excited state was taken from predictions from the continuum shell-model~\cite{Volya06}.

\par
\begin{figure}
\includegraphics[width=0.4\textwidth]{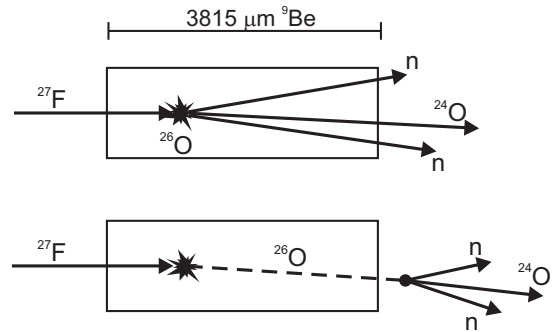}
\caption{\label{f:cartoon}  The decay of $^{26}$O within the thick $^{9}$Be target is illustrated for two cases: (top) very short lifetime corresponding to an immediate decay and (bottom) a lifetime around 30~ps which allows the $^{26}$O to exit the target before decaying.}
\end{figure}

Two scenarios for the decay of $^{26}$O with different lifetimes are illustrated in Fig.~\ref{f:cartoon}.  The $^{27}$F beam enters the 3815 $\mu$m (705 mg/cm$^{2}$) $^{9}$Be target followed by the one-proton knockout reaction producing the $^{26}$O.  If the reaction was to occur at the beginning of the target where the $^{27}$F beam is traveling at about 11.8 cm/ns and the $^{26}$O had a very short lifetime (top of Fig.~\ref{f:cartoon}) then the neutrons would be emitted with an average velocity of 11.8 cm/ns.  In the other case if the $^{26}$O had a lifetime of 30~ps (bottom of Fig.~\ref{f:cartoon}), which is roughly the time of flight through the target, then the neutrons would be emitted with an average velocity of 10.9 cm/ns due to the energy loss of the $^{26}$O fragment traveling through the target.  Thus, the observation of a shift in the expected neutron velocity can provide a measure of the lifetime of $^{26}$O.

%The kinetic energy (and velocity) of the fragment after exiting the target is determined from the track of the fragment through the 4~Tm dipole magnet.

\par
The relative velocity between the neutrons and fragment is defined as $V_{rel} = V_{n} - V_{frag}$, where $V_{n}$ ($V_{frag}$) is the velocity of the neutron (fragment) in the laboratory frame.  The relative velocity was examined to remove the effect of the momentum dispersion of the $^{27}$F beam ($\Delta p/p$~=~2$\%$).  Thus, the variation in the incoming velocity of the $^{27}$F is removed event by event. Since the reaction point in the target is unknown, the fragment velocity ($V_{frag}$) is calculated assuming the reaction occurs at the center of the target.  The width of the $V_{rel}$ distribution will be dependent on the target thickness and magnitude of the decay energy (both of which will increase the width).  If the reaction point was known on an event-by-event basis and the decay energy was very small then the $V_{rel}$ distribution should be narrow and centered around zero.  While the width of the experimental $V_{rel}$ distribution will be increased, the mean velocity will still be centered around zero if the $^{26}$O lifetime is short.  Thus, a shift in the relative velocity away from zero would indicate a long-lived component of the decay.

\par
\begin{figure}
\includegraphics[width=0.38\textwidth]{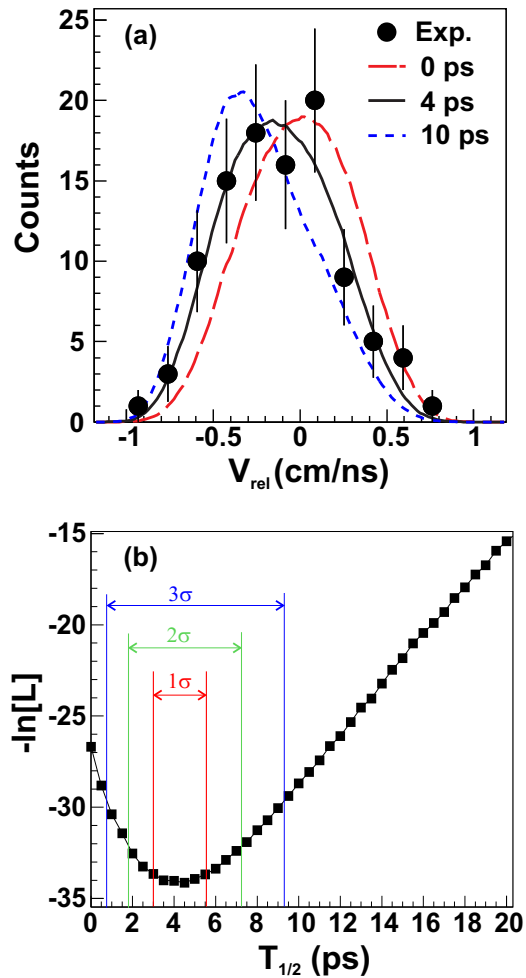}
\caption{\label{f:vrel} (Color online) (a) Experimental $V_{rel}$ distribution from the decay of $^{26}$O compared to the Monte Carlo simulation where the $^{26}$O half-life is set as 0~ps, 4~ps, and 10~ps. (b) Negative log-likelihood ($-$ln$[L]$) as a function of the half-life from the unbinned fit of the experimental $V_{rel}$ data set.  The 1$\sigma$, 2$\sigma$, and 3$\sigma$ confidence interval are indicated.}
\end{figure}

The $^{26}$O events were selected from events passing the causality cut criteria and having a $E_{decay}<1.0$~MeV, as indicated from the dotted grey line in Fig.~\ref{f:edecay}.  This selection should maximize the statistics and minimize the contamination of other decay channels.  Based on the fit of Fig.~\ref{f:edecay} the $^{26}$O ground state resonance accounts for 96$\%$ of the events with $E_{decay}<1.0$~MeV. The relative velocity between the $^{24}$O and each of the emitted neutrons is shown in Fig.~\ref{f:vrel}(a). The experimental $V_{rel}$ (solid black points) is shifted away from zero with an average $V_{rel}<0$.  This implies that the neutron velocity is smaller than the fragment velocity (at half target thickness).  This would be the case if the neutrons were emitted after the $^{26}$O traveled through a portion of the target decreasing its velocity.

It is important to understand the calibration, resolution, and accuracy of the neutron and fragment velocities. The neutron time-of-flight (and therefore velocity) was calibrated based on the time-of-flight of the gamma-rays produced at the target traveling to MoNA.  From the width of the gamma peak the relative resolution (FWHM/centroid) of the neutron velocity is about 3$\%$.  In comparison to the resolution, the accuracy of the neutron velocity is determined from the accuracy of the time-of-flight measurement and location of the MoNA detector. The neutron velocity accuracy was determined to be 0.03 cm/ns at beam velocity (11.8 cm/ns).  The fragment velocity is determined from the track of the fragment through the dipole magnet which is measured using two Cathode Readout Drift Chambers (CRDCs).  The accuracy of the fragment velocity is related to the accuracy of the magnetic field map and measured position of the CRDCs. Reasonable variation of these parameters showed the fragment velocity to be 0.02 cm/ns at beam velocity. While the resolutions will determine the width of the $V_{rel}$ distribution, the accuracy of the centroid is related to the accuracy of the neutron and fragment velocity measurements. The detector resolutions were included in the Monte Carlo simulation.

\par
In order to extract a half-life limit ($T_{1/2}$) of $^{26}$O from the $V_{rel}$ distribution, the Monte Carlo simulation was modified such that the probability distribution for the $^{26}$O decay based on $T_{1/2}$ was included.  Thus, after the one-proton reaction occurred within the target (at a random position) the $^{26}$O was propagated for a time $t$ determined from the probability distribution before decaying into $^{24}\mathrm{O}+n+n$.  The resulting $V_{rel}$ distributions from the simulation with $T_{1/2}$~=~0, 4, and 10~ps is compared to the experiment in Fig.~\ref{f:vrel}(a).  The $V_{rel}$ distribution with $T_{1/2}$~=~0 is unable to reproduce the experimental data.  A much better fit is achieved with $T_{1/2}$~=~4~ps, which shows a similar shift in the simulation as the experiment.  This suggests that $^{26}$O did not decay instantaneously but had an appreciable lifetime.  The shape of the decay energy spectrum (Fig.~\ref{f:edecay}) would also be affected by the finite lifetime of $^{26}$O.  The Monte Carlo simulations showed that significant changes in the $E_{decay}$ spectrum would be observed for $T_{1/2} \gtrsim$~10~ps.

\par
Due to the low statistics of the experiment, the $\chi^{2}$ analysis was observed to be dependent on the binning of the data.  Therefore, a unbinned maximum likelihood technique was employed to determine the statistical significance of the results.  This procedure is described in Ref.~\cite{Sch93} and was recently used in the analysis of $^{27,28}$F measured with MoNA~\cite{Chr12,Chr12PRL}.  The negative log-likelihood ($-$ln$[L]$) is plotted as function of $T_{1/2}$ in Fig.~\ref{f:vrel}(b).  A minimum in $-$ln$[L]$ is found at 4.5~ps.  The $n\sigma$ confidence intervals are calculated as ln$[L_{max}] - $ln$[L]\leq n^{2}/2$.  The 1$\sigma$, 2$\sigma$, and 3$\sigma$ confidence intervals are shown in Fig.~\ref{f:vrel}(b).  The statistical significance of the results indicate that $^{26}$O has a half-life of about 4.5~ps.

\par
\begin{figure}
\includegraphics[width=0.30\textwidth]{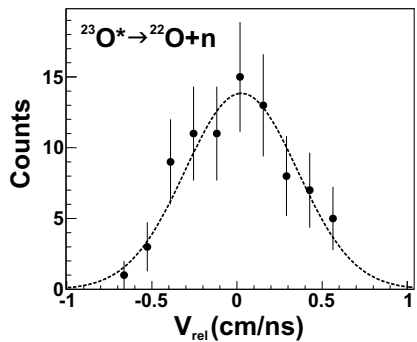}
\caption{\label{f:o23} Experimental $V_{rel}$ distribution from the $^{23}$O$^{*} \rightarrow ^{22}$O$ + n$ decay.  The Gaussian fit (dashed line) is shown to guide the eye.  }
\end{figure}

In addition to the statistical significance, it is important to account for possible systematic uncertainties.  As previously discussed, the accuracy of the neutron and fragment velocities was 0.03 and 0.02 cm/ns, respectively.  This represents a total systematic uncertainty of 0.05 cm/ns in the $V_{rel}$ distribution, which corresponds to a 1.7~ps systematic uncertainty, and indicates a finite half-life of $^{26}$O at 95$\%$ confidence level. The systematic uncertainty was also estimated through examining the neutron decay of the first excited state of $^{23}$O$^{*} \rightarrow ^{22}$O$ + n$, which was also populated during the experiment from the $^{27}$F beam.  Thus, the $^{9}$Be target, $B\rho$ of the dipole, MoNA configuration, Sweeper detector settings, and calibrations were identical to the $^{26}$O measurement and can provide an estimate of any unknown systematic errors.  Since the decay of $^{23}$O should not have a long-lived component, the relative velocity spectrum should not be shifted away from $V_{rel}$~=~0 (see Fig.~\ref{f:o23}).  Following the same half-life analysis discussed above, the upper limit on $1\sigma$ $T_{1/2}$ was 3~ps for the $^{23}$O distribution.  Therefore, the systematic uncertainty was estimated as 3~ps in comparison to 1.7~ps determined above.  The half-life of $^{26}$O is then taken as 4.5$^{+1.1}_{-1.5}~(stat.) \pm 3 (sys.)$~ps, which gives $T_{1/2} > 0$ at 82$\%$ confidence level.  A new measurement with improved statistics would allow for both the statistical and systematic uncertainties to be reduced.

\begin{figure}
\includegraphics[width=0.41\textwidth]{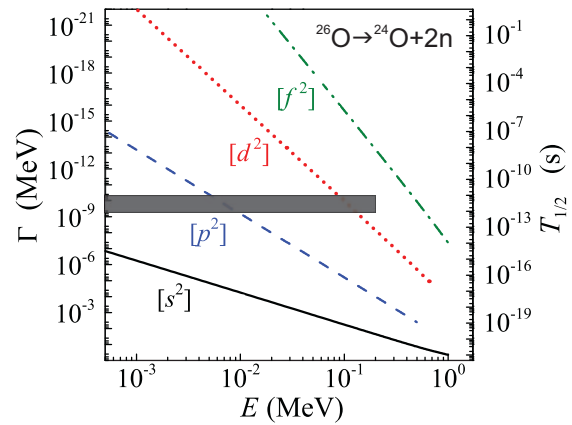}
\caption{\label{f:grig} (Color online) Predications of Grigorenko \emph{et al.} for the true two-neutron emission width (half-life) as a function of the decay energy.  The grey region represents the $3\sigma$ statistical limit on the $T_{1/2}$ from the current work and the ground state resonance energy limit from Ref.~\cite{LUN12}.  This figure was adapted from Ref.~\cite{Gri11}.  }
\end{figure}
% 2n, 1s is 2.95 to 5.7, 2s is 1.85 to 7.35 and 3s is 0.8 to 9.3

\par
Possible systematic effects related to the selection of the $^{24}$O events and the application of the $2n$ causality cuts were also investigated.  The selection of the $^{24}$O events (shown in Fig. 2 of Ref.~\cite{LUN12}) based on the time-of-flight was varied to examine the dependence of the $V_{rel}$ distribution (or half-life).  The results showed minor variations and the half-life remained within the 1$\sigma$ limit shown in Fig.~\ref{f:vrel}.  While the causality cuts allow for the removal of the large majority of all false $2n$ events, it is important to verify that shift in the $V_{rel}$ distribution is not created by the causality cuts.  The causality cuts were removed from both the experimental and simulated data and the shift in the $V_{rel}$ distribution was maintained.

\par
The results of the lifetime analysis are compared with the prediction of Grigorenko \emph{et al.}~\cite{Gri11} in Fig.~\ref{f:grig}.  The solid grey region is defined from the $3\sigma$ statistical limit on the $T_{1/2}$ and the ground state resonance energy limit of $E<200$~keV~\cite{LUN12} for $^{26}$O.  As shown the agreement with the predictions will depend greatly on improving the constraints on the energy and the configuration of the ground state.  For example, a small $\nu(s)^{2}$ component in the $^{26}$O ground state will require the resonance energy to be very small for $T_{1/2}$~=~4.5~ps.

\par
In summary, a new technique for measuring the lifetimes of neutron unbound nuclei has been presented and applied to the case of $^{26}$O.  A shift in the relative velocity between the $^{24}$O and emitted neutrons was observed and shown to be related to the lifetime of $^{26}$O.  Detailed Monte Carlo simulations, in which the half-life of $^{26}$O could be varied, were compared to the experimental data.  The extracted $^{26}$O half-life was 4.5$^{+1.1}_{-1.5}~(stat.) \pm 3 (sys.)$~ps.  This corresponds to $^{26}$O having a finite lifetime with an 82$\%$ confidence level and suggests the possibility of two-neutron radioactivity.  It appears that, much like the proton dripline, the unique structure of the neutron dripline nuclei opens the door for observations of new modes of radioactivity.  Future experimental work is needed to confirm this observation and provide stringent constraints on the properties of the $^{26}$O ground state.

\par
The authors would like to thank H. Attanayake, D. Divaratne, S. M. Grimes, A. Haagsma, and A. Schiller from Ohio University for help during the experiment. The authors gratefully acknowledge the support of the NSCL operations staff for providing a high quality beam.  This material is based upon work supported by the Department of Energy National Nuclear Security Administration under Award Number DE-NA0000979 and DOE Award number DE-FG02-92ER40750.  This work was also supported by the National Science Foundation under Grant Nos. PHY06-06007, PHY08-55456, PHY09-22335, PHY09-69058, and PHY11-02511.

%\clearpage %Just because of unusual number of tables stacked at end
%\bibliography{bibdata}% Produces the bibliography via BibTeX.

\begin{thebibliography}{33}%
\makeatletter
\providecommand \@ifxundefined [1]{%
 \@ifx{#1\undefined}
}%
\providecommand \@ifnum [1]{%
 \ifnum #1\expandafter \@firstoftwo
 \else \expandafter \@secondoftwo
 \fi
}%
\providecommand \@ifx [1]{%
 \ifx #1\expandafter \@firstoftwo
 \else \expandafter \@secondoftwo
 \fi
}%
\providecommand \natexlab [1]{#1}%
\providecommand \enquote  [1]{``#1''}%
\providecommand \bibnamefont  [1]{#1}%
\providecommand \bibfnamefont [1]{#1}%
\providecommand \citenamefont [1]{#1}%
\providecommand \href@noop [0]{\@secondoftwo}%
\providecommand \href [0]{\begingroup \@sanitize@url \@href}%
\providecommand \@href[1]{\@@startlink{#1}\@@href}%
\providecommand \@@href[1]{\endgroup#1\@@endlink}%
\providecommand \@sanitize@url [0]{\catcode `\\12\catcode `\$12\catcode
  `\&12\catcode `\#12\catcode `\^12\catcode `\_12\catcode `\%12\relax}%
\providecommand \@@startlink[1]{}%
\providecommand \@@endlink[0]{}%
\providecommand \url  [0]{\begingroup\@sanitize@url \@url }%
\providecommand \@url [1]{\endgroup\@href {#1}{\urlprefix }}%
\providecommand \urlprefix  [0]{URL }%
\providecommand \Eprint [0]{\href }%
\providecommand \doibase [0]{http://dx.doi.org/}%
\providecommand \selectlanguage [0]{\@gobble}%
\providecommand \bibinfo  [0]{\@secondoftwo}%
\providecommand \bibfield  [0]{\@secondoftwo}%
\providecommand \translation [1]{[#1]}%
\providecommand \BibitemOpen [0]{}%
\providecommand \bibitemStop [0]{}%
\providecommand \bibitemNoStop [0]{.\EOS\space}%
\providecommand \EOS [0]{\spacefactor3000\relax}%
\providecommand \BibitemShut  [1]{\csname bibitem#1\endcsname}%
\let\auto@bib@innerbib\@empty
%</preamble>
\bibitem [{\citenamefont {Baumann}\ \emph {et~al.}(2012)\citenamefont
  {Baumann}, \citenamefont {Spyrou},\ and\ \citenamefont
  {Thoennessen}}]{BAUMANN12}%
  \BibitemOpen
  \bibfield  {author} {\bibinfo {author} {\bibfnamefont {T.}~\bibnamefont
  {Baumann}}, \bibinfo {author} {\bibfnamefont {A.}~\bibnamefont {Spyrou}}, \
  and\ \bibinfo {author} {\bibfnamefont {M.}~\bibnamefont {Thoennessen}},\
  }\href@noop {} {\bibfield  {journal} {\bibinfo  {journal} {Rep. Prog. Phys.}\
  }\textbf {\bibinfo {volume} {75}},\ \bibinfo {pages} {036301} (\bibinfo
  {year} {2012})}\BibitemShut {NoStop}%
\bibitem [{\citenamefont {Pfutzner}\ \emph {et~al.}(2012)\citenamefont
  {Pfutzner}, \citenamefont {Karny}, \citenamefont {Grigorenko},\ and\
  \citenamefont {Rissager}}]{Pfu12}%
  \BibitemOpen
  \bibfield  {author} {\bibinfo {author} {\bibfnamefont {M.}~\bibnamefont
  {Pfutzner}}, \bibinfo {author} {\bibfnamefont {M.}~\bibnamefont {Karny}},
  \bibinfo {author} {\bibfnamefont {L.~V.}\ \bibnamefont {Grigorenko}}, \ and\
  \bibinfo {author} {\bibfnamefont {K.}~\bibnamefont {Rissager}},\ }\href@noop
  {} {\bibfield  {journal} {\bibinfo  {journal} {Rev. Mod. Phys.}\ }\textbf
  {\bibinfo {volume} {84}},\ \bibinfo {pages} {567} (\bibinfo {year}
  {2012})}\BibitemShut {NoStop}%
\bibitem [{\citenamefont {Thoennessen}(2004)}]{Tho04}%
  \BibitemOpen
  \bibfield  {author} {\bibinfo {author} {\bibfnamefont {M.}~\bibnamefont
  {Thoennessen}},\ }\href@noop {} {\bibfield  {journal} {\bibinfo  {journal}
  {Rep. Prog. Phys.}\ }\textbf {\bibinfo {volume} {67}},\ \bibinfo {pages}
  {1187} (\bibinfo {year} {2004})}\BibitemShut {NoStop}%
\bibitem [{\citenamefont {Hansen}\ and\ \citenamefont
  {Sherrill}(2001)}]{Han01}%
  \BibitemOpen
  \bibfield  {author} {\bibinfo {author} {\bibfnamefont {P.~G.}\ \bibnamefont
  {Hansen}}\ and\ \bibinfo {author} {\bibfnamefont {B.~M.}\ \bibnamefont
  {Sherrill}},\ }\href@noop {} {\bibfield  {journal} {\bibinfo  {journal}
  {Nucl. Phys. A}\ }\textbf {\bibinfo {volume} {693}},\ \bibinfo {pages} {133}
  (\bibinfo {year} {2001})}\BibitemShut {NoStop}%
\bibitem [{\citenamefont {Brown}(2001)}]{BROWN01}%
  \BibitemOpen
  \bibfield  {author} {\bibinfo {author} {\bibfnamefont {B.~A.}\ \bibnamefont
  {Brown}},\ }\href@noop {} {\bibfield  {journal} {\bibinfo  {journal} {Prog.
  Part. Nucl. Phys.}\ }\textbf {\bibinfo {volume} {47}},\ \bibinfo {pages}
  {517} (\bibinfo {year} {2001})}\BibitemShut {NoStop}%
\bibitem [{\citenamefont {Otsuka}\ \emph {et~al.}(2001)\citenamefont {Otsuka},
  \citenamefont {Honma}, \citenamefont {Mizusaki}, \citenamefont {Shimizu},\
  and\ \citenamefont {Utsuno}}]{Ots01}%
  \BibitemOpen
  \bibfield  {author} {\bibinfo {author} {\bibfnamefont {T.}~\bibnamefont
  {Otsuka}}, \bibinfo {author} {\bibfnamefont {M.}~\bibnamefont {Honma}},
  \bibinfo {author} {\bibfnamefont {T.}~\bibnamefont {Mizusaki}}, \bibinfo
  {author} {\bibfnamefont {N.}~\bibnamefont {Shimizu}}, \ and\ \bibinfo
  {author} {\bibfnamefont {Y.}~\bibnamefont {Utsuno}},\ }\href@noop {}
  {\bibfield  {journal} {\bibinfo  {journal} {Prog. Part. Nucl. Phys.}\
  }\textbf {\bibinfo {volume} {47}},\ \bibinfo {pages} {319.} (\bibinfo {year}
  {2001})}\BibitemShut {NoStop}%
\bibitem [{\citenamefont {Goldansky}(1960)}]{Gold60}%
  \BibitemOpen
  \bibfield  {author} {\bibinfo {author} {\bibfnamefont {V.~I.}\ \bibnamefont
  {Goldansky}},\ }\href@noop {} {\bibfield  {journal} {\bibinfo  {journal}
  {Nucl. Phys.}\ }\textbf {\bibinfo {volume} {19}},\ \bibinfo {pages} {482}
  (\bibinfo {year} {1960})}\BibitemShut {NoStop}%
\bibitem [{\citenamefont {Hofmann}\ \emph {et~al.}(1982)\citenamefont
  {Hofmann}, \citenamefont {Reisdorf}, \citenamefont {Munzenberg},
  \citenamefont {He$\ss$berger}, \citenamefont {Schneider},\ and\ \citenamefont
  {Armbruster}}]{Hof82}%
  \BibitemOpen
  \bibfield  {author} {\bibinfo {author} {\bibfnamefont {S.}~\bibnamefont
  {Hofmann}}, \bibinfo {author} {\bibfnamefont {W.}~\bibnamefont {Reisdorf}},
  \bibinfo {author} {\bibfnamefont {G.}~\bibnamefont {Munzenberg}}, \bibinfo
  {author} {\bibfnamefont {F.~P.}\ \bibnamefont {He$\ss$berger}}, \bibinfo
  {author} {\bibfnamefont {J.~R.~H.}\ \bibnamefont {Schneider}}, \ and\
  \bibinfo {author} {\bibfnamefont {P.}~\bibnamefont {Armbruster}},\
  }\href@noop {} {\bibfield  {journal} {\bibinfo  {journal} {Z. Phys. A}\
  }\textbf {\bibinfo {volume} {305}},\ \bibinfo {pages} {111} (\bibinfo {year}
  {1982})}\BibitemShut {NoStop}%
\bibitem [{\citenamefont {Klepper}\ \emph {et~al.}(1982)\citenamefont
  {Klepper}, \citenamefont {Batsch}, \citenamefont {Hofmann}, \citenamefont
  {Kirchner}, \citenamefont {kurcewicz}, \citenamefont {Residorf},
  \citenamefont {Roeckl}, \citenamefont {Schardt},\ and\ \citenamefont
  {Nyman}}]{Klep82}%
  \BibitemOpen
  \bibfield  {author} {\bibinfo {author} {\bibfnamefont {O.}~\bibnamefont
  {Klepper}}, \bibinfo {author} {\bibfnamefont {T.}~\bibnamefont {Batsch}},
  \bibinfo {author} {\bibfnamefont {S.}~\bibnamefont {Hofmann}}, \bibinfo
  {author} {\bibfnamefont {R.}~\bibnamefont {Kirchner}}, \bibinfo {author}
  {\bibfnamefont {W.}~\bibnamefont {kurcewicz}}, \bibinfo {author}
  {\bibfnamefont {W.}~\bibnamefont {Residorf}}, \bibinfo {author}
  {\bibfnamefont {E.}~\bibnamefont {Roeckl}}, \bibinfo {author} {\bibfnamefont
  {D.}~\bibnamefont {Schardt}}, \ and\ \bibinfo {author} {\bibfnamefont
  {G.}~\bibnamefont {Nyman}},\ }\href@noop {} {\bibfield  {journal} {\bibinfo
  {journal} {Z. Phys. A}\ }\textbf {\bibinfo {volume} {305}},\ \bibinfo {pages}
  {125} (\bibinfo {year} {1982})}\BibitemShut {NoStop}%
\bibitem [{\citenamefont {Pfutzner}\ \emph {et~al.}(2002)\citenamefont
  {Pfutzner} \emph {et~al.}}]{Pfu02}%
  \BibitemOpen
  \bibfield  {author} {\bibinfo {author} {\bibfnamefont {M.}~\bibnamefont
  {Pfutzner}} \emph {et~al.},\ }\href@noop {} {\bibfield  {journal} {\bibinfo
  {journal} {Eur. Phys. J. A}\ }\textbf {\bibinfo {volume} {14}},\ \bibinfo
  {pages} {279} (\bibinfo {year} {2002})}\BibitemShut {NoStop}%
\bibitem [{\citenamefont {Giovinazzo}\ \emph {et~al.}(2002)\citenamefont
  {Giovinazzo} \emph {et~al.}}]{Gio02}%
  \BibitemOpen
  \bibfield  {author} {\bibinfo {author} {\bibfnamefont {J.}~\bibnamefont
  {Giovinazzo}} \emph {et~al.},\ }\href@noop {} {\bibfield  {journal} {\bibinfo
   {journal} {Phys. Rev. Lett.}\ }\textbf {\bibinfo {volume} {89}},\ \bibinfo
  {pages} {102501} (\bibinfo {year} {2002})}\BibitemShut {NoStop}%
\bibitem [{\citenamefont {Miernik}\ \emph {et~al.}(2007)\citenamefont {Miernik}
  \emph {et~al.}}]{Mie07}%
  \BibitemOpen
  \bibfield  {author} {\bibinfo {author} {\bibfnamefont {K.}~\bibnamefont
  {Miernik}} \emph {et~al.},\ }\href@noop {} {\bibfield  {journal} {\bibinfo
  {journal} {Phys. Rev. Lett.}\ }\textbf {\bibinfo {volume} {99}},\ \bibinfo
  {pages} {192501} (\bibinfo {year} {2007})}\BibitemShut {NoStop}%
\bibitem [{\citenamefont {Azuma}\ \emph {et~al.}(1979)\citenamefont {Azuma},
  \citenamefont {Carraz}, \citenamefont {Hansen}, \citenamefont {Jonson},
  \citenamefont {Kratz}, \citenamefont {Mattsson}, \citenamefont {Nyman},
  \citenamefont {Ohm}, \citenamefont {Ravn}, \citenamefont {Schroder},\ and\
  \citenamefont {Ziegert}}]{Asu79}%
  \BibitemOpen
  \bibfield  {author} {\bibinfo {author} {\bibfnamefont {R.~E.}\ \bibnamefont
  {Azuma}}, \bibinfo {author} {\bibfnamefont {L.~C.}\ \bibnamefont {Carraz}},
  \bibinfo {author} {\bibfnamefont {P.~G.}\ \bibnamefont {Hansen}}, \bibinfo
  {author} {\bibfnamefont {B.}~\bibnamefont {Jonson}}, \bibinfo {author}
  {\bibfnamefont {K.~L.}\ \bibnamefont {Kratz}}, \bibinfo {author}
  {\bibfnamefont {S.}~\bibnamefont {Mattsson}}, \bibinfo {author}
  {\bibfnamefont {G.}~\bibnamefont {Nyman}}, \bibinfo {author} {\bibfnamefont
  {H.}~\bibnamefont {Ohm}}, \bibinfo {author} {\bibfnamefont {H.~L.}\
  \bibnamefont {Ravn}}, \bibinfo {author} {\bibfnamefont {A.}~\bibnamefont
  {Schroder}}, \ and\ \bibinfo {author} {\bibfnamefont {W.}~\bibnamefont
  {Ziegert}},\ }\href@noop {} {\bibfield  {journal} {\bibinfo  {journal} {Phys.
  Rev. Lett.}\ }\textbf {\bibinfo {volume} {43}},\ \bibinfo {pages} {1652}
  (\bibinfo {year} {1979})}\BibitemShut {NoStop}%
\bibitem [{\citenamefont {Azuma}\ \emph {et~al.}(1980)\citenamefont {Azuma},
  \citenamefont {Bjornstad}, \citenamefont {Gustafsson}, \citenamefont
  {Hansen}, \citenamefont {Johnson}, \citenamefont {Mattsson}, \citenamefont
  {Nyman}, \citenamefont {Poskanzer},\ and\ \citenamefont {Ravn}}]{Azu80}%
  \BibitemOpen
  \bibfield  {author} {\bibinfo {author} {\bibfnamefont {R.~E.}\ \bibnamefont
  {Azuma}}, \bibinfo {author} {\bibfnamefont {T.}~\bibnamefont {Bjornstad}},
  \bibinfo {author} {\bibfnamefont {H.~A.}\ \bibnamefont {Gustafsson}},
  \bibinfo {author} {\bibfnamefont {P.~G.}\ \bibnamefont {Hansen}}, \bibinfo
  {author} {\bibfnamefont {B.}~\bibnamefont {Johnson}}, \bibinfo {author}
  {\bibfnamefont {S.}~\bibnamefont {Mattsson}}, \bibinfo {author}
  {\bibfnamefont {G.}~\bibnamefont {Nyman}}, \bibinfo {author} {\bibfnamefont
  {A.~M.}\ \bibnamefont {Poskanzer}}, \ and\ \bibinfo {author} {\bibfnamefont
  {H.~L.}\ \bibnamefont {Ravn}},\ }\href@noop {} {\bibfield  {journal}
  {\bibinfo  {journal} {Phys. Lett. B}\ }\textbf {\bibinfo {volume} {96}},\
  \bibinfo {pages} {31} (\bibinfo {year} {1980})}\BibitemShut {NoStop}%
\bibitem [{\citenamefont {Dufour}\ \emph {et~al.}(1988)\citenamefont {Dufour}
  \emph {et~al.}}]{Duf88}%
  \BibitemOpen
  \bibfield  {author} {\bibinfo {author} {\bibfnamefont {J.~P.}\ \bibnamefont
  {Dufour}} \emph {et~al.},\ }\href@noop {} {\bibfield  {journal} {\bibinfo
  {journal} {Phys. Lett. B}\ }\textbf {\bibinfo {volume} {206}},\ \bibinfo
  {pages} {195} (\bibinfo {year} {1988})}\BibitemShut {NoStop}%
\bibitem [{\citenamefont {Grigorenko}\ \emph {et~al.}(2011)\citenamefont
  {Grigorenko}, \citenamefont {Mukha}, \citenamefont {Scheidenberger},\ and\
  \citenamefont {Zhukov}}]{Gri11}%
  \BibitemOpen
  \bibfield  {author} {\bibinfo {author} {\bibfnamefont {L.~V.}\ \bibnamefont
  {Grigorenko}}, \bibinfo {author} {\bibfnamefont {I.~G.}\ \bibnamefont
  {Mukha}}, \bibinfo {author} {\bibfnamefont {C.}~\bibnamefont
  {Scheidenberger}}, \ and\ \bibinfo {author} {\bibfnamefont {M.~V.}\
  \bibnamefont {Zhukov}},\ }\href@noop {} {\bibfield  {journal} {\bibinfo
  {journal} {Phys. Rev. C}\ }\textbf {\bibinfo {volume} {84}},\ \bibinfo
  {pages} {021303(R)} (\bibinfo {year} {2011})}\BibitemShut {NoStop}%
\bibitem [{\citenamefont {Lunderberg}\ \emph {et~al.}(2012)\citenamefont
  {Lunderberg}, \citenamefont {DeYoung}, \citenamefont {Kohley}, \citenamefont
  {Attanayake}, \citenamefont {Baumann}, \citenamefont {Bazin}, \citenamefont
  {Christian}, \citenamefont {Divaratne}, \citenamefont {Grimes}, \citenamefont
  {Haagsma}, \citenamefont {Finck}, \citenamefont {Frank} \emph
  {et~al.}}]{LUN12}%
  \BibitemOpen
  \bibfield  {author} {\bibinfo {author} {\bibfnamefont {E.}~\bibnamefont
  {Lunderberg}}, \bibinfo {author} {\bibfnamefont {P.~A.}\ \bibnamefont
  {DeYoung}}, \bibinfo {author} {\bibfnamefont {Z.}~\bibnamefont {Kohley}},
  \bibinfo {author} {\bibfnamefont {H.}~\bibnamefont {Attanayake}}, \bibinfo
  {author} {\bibfnamefont {T.}~\bibnamefont {Baumann}}, \bibinfo {author}
  {\bibfnamefont {D.}~\bibnamefont {Bazin}}, \bibinfo {author} {\bibfnamefont
  {G.}~\bibnamefont {Christian}}, \bibinfo {author} {\bibfnamefont
  {D.}~\bibnamefont {Divaratne}}, \bibinfo {author} {\bibfnamefont {S.~M.}\
  \bibnamefont {Grimes}}, \bibinfo {author} {\bibfnamefont {A.}~\bibnamefont
  {Haagsma}}, \bibinfo {author} {\bibfnamefont {J.~E.}\ \bibnamefont {Finck}},
  \bibinfo {author} {\bibfnamefont {N.}~\bibnamefont {Frank}},  \emph
  {et~al.},\ }\href@noop {} {\bibfield  {journal} {\bibinfo  {journal} {Phys.
  Rev. Lett.}\ }\textbf {\bibinfo {volume} {108}},\ \bibinfo {pages} {142503}
  (\bibinfo {year} {2012})}\BibitemShut {NoStop}%
\bibitem [{\citenamefont {Caesar}\ \emph {et~al.}(2012)\citenamefont {Caesar}
  \emph {et~al.}}]{Cae12}%
  \BibitemOpen
  \bibfield  {author} {\bibinfo {author} {\bibfnamefont {C.}~\bibnamefont
  {Caesar}} \emph {et~al.},\ }\href@noop {} {\bibfield  {journal} {\bibinfo
  {journal} {arXiv:1209.0156 [nucl-ex]}\ } (\bibinfo {year}
  {2012})}\BibitemShut {NoStop}%
\bibitem [{\citenamefont {Guillemaud-Mueller}\ \emph
  {et~al.}(1990)\citenamefont {Guillemaud-Mueller} \emph {et~al.}}]{Gui90}%
  \BibitemOpen
  \bibfield  {author} {\bibinfo {author} {\bibfnamefont {D.}~\bibnamefont
  {Guillemaud-Mueller}} \emph {et~al.},\ }\href@noop {} {\bibfield  {journal}
  {\bibinfo  {journal} {Phys. Rev. C}\ }\textbf {\bibinfo {volume} {41}},\
  \bibinfo {pages} {937} (\bibinfo {year} {1990})}\BibitemShut {NoStop}%
\bibitem [{\citenamefont {Tarasov}\ \emph {et~al.}(1997)\citenamefont {Tarasov}
  \emph {et~al.}}]{Tar97}%
  \BibitemOpen
  \bibfield  {author} {\bibinfo {author} {\bibfnamefont {O.}~\bibnamefont
  {Tarasov}} \emph {et~al.},\ }\href@noop {} {\bibfield  {journal} {\bibinfo
  {journal} {Phys. Lett. B}\ }\textbf {\bibinfo {volume} {409}},\ \bibinfo
  {pages} {64} (\bibinfo {year} {1997})}\BibitemShut {NoStop}%
\bibitem [{\citenamefont {Liddick}\ \emph {et~al.}(2006)\citenamefont {Liddick}
  \emph {et~al.}}]{Lid06}%
  \BibitemOpen
  \bibfield  {author} {\bibinfo {author} {\bibfnamefont {S.~N.}\ \bibnamefont
  {Liddick}} \emph {et~al.},\ }\href@noop {} {\bibfield  {journal} {\bibinfo
  {journal} {Phys. Rev. Lett.}\ }\textbf {\bibinfo {volume} {97}},\ \bibinfo
  {pages} {082501} (\bibinfo {year} {2006})}\BibitemShut {NoStop}%
\bibitem [{\citenamefont {Mukha}\ \emph {et~al.}(2007)\citenamefont {Mukha}
  \emph {et~al.}}]{Muk07}%
  \BibitemOpen
  \bibfield  {author} {\bibinfo {author} {\bibfnamefont {I.}~\bibnamefont
  {Mukha}} \emph {et~al.},\ }\href@noop {} {\bibfield  {journal} {\bibinfo
  {journal} {Phys. Rev. Lett.}\ }\textbf {\bibinfo {volume} {99}},\ \bibinfo
  {pages} {182501} (\bibinfo {year} {2007})}\BibitemShut {NoStop}%
\bibitem [{\citenamefont {Voss}\ \emph {et~al.}(2012)\citenamefont {Voss} \emph
  {et~al.}}]{Voss12}%
  \BibitemOpen
  \bibfield  {author} {\bibinfo {author} {\bibfnamefont {P.~J.}\ \bibnamefont
  {Voss}} \emph {et~al.},\ }\href@noop {} {\enquote {\bibinfo {title}
  {Two-proton decay lifetime 19mg},}\ } (\bibinfo {year} {2012}),\ \bibinfo
  {note} {submitted to Phys. Rev. C}\BibitemShut {NoStop}%
\bibitem [{\citenamefont {Schwarzschild}\ and\ \citenamefont
  {Warburton}(1968)}]{Sch68}%
  \BibitemOpen
  \bibfield  {author} {\bibinfo {author} {\bibfnamefont {A.~Z.}\ \bibnamefont
  {Schwarzschild}}\ and\ \bibinfo {author} {\bibfnamefont {W.~K.}\ \bibnamefont
  {Warburton}},\ }\href@noop {} {\bibfield  {journal} {\bibinfo  {journal}
  {Annu. Rev. Nucl. Sci.}\ }\textbf {\bibinfo {volume} {18}},\ \bibinfo {pages}
  {265} (\bibinfo {year} {1968})}\BibitemShut {NoStop}%
\bibitem [{\citenamefont {Nolan}\ and\ \citenamefont
  {Sharpey-Schafer}(1979)}]{Nol79}%
  \BibitemOpen
  \bibfield  {author} {\bibinfo {author} {\bibfnamefont {P.~J.}\ \bibnamefont
  {Nolan}}\ and\ \bibinfo {author} {\bibfnamefont {J.~F.}\ \bibnamefont
  {Sharpey-Schafer}},\ }\href@noop {} {\bibfield  {journal} {\bibinfo
  {journal} {Rep. Prog. Phys.}\ }\textbf {\bibinfo {volume} {42}},\ \bibinfo
  {pages} {1} (\bibinfo {year} {1979})}\BibitemShut {NoStop}%
\bibitem [{\citenamefont {Dewald}\ \emph {et~al.}(2012)\citenamefont {Dewald},
  \citenamefont {Moller},\ and\ \citenamefont {Petkov}}]{Dew12}%
  \BibitemOpen
  \bibfield  {author} {\bibinfo {author} {\bibfnamefont {A.}~\bibnamefont
  {Dewald}}, \bibinfo {author} {\bibfnamefont {O.}~\bibnamefont {Moller}}, \
  and\ \bibinfo {author} {\bibfnamefont {P.}~\bibnamefont {Petkov}},\
  }\href@noop {} {\bibfield  {journal} {\bibinfo  {journal} {Prog. Part. Nucl.
  Phys.}\ }\textbf {\bibinfo {volume} {67}},\ \bibinfo {pages} {786} (\bibinfo
  {year} {2012})}\BibitemShut {NoStop}%
\bibitem [{\citenamefont {Morrissey}\ \emph {et~al.}(2003)\citenamefont
  {Morrissey}, \citenamefont {Sherrill}, \citenamefont {Steiner}, \citenamefont
  {Stolz},\ and\ \citenamefont {Wiedenhoever}}]{Mor03}%
  \BibitemOpen
  \bibfield  {author} {\bibinfo {author} {\bibfnamefont {D.~J.}\ \bibnamefont
  {Morrissey}}, \bibinfo {author} {\bibfnamefont {B.~M.}\ \bibnamefont
  {Sherrill}}, \bibinfo {author} {\bibfnamefont {M.}~\bibnamefont {Steiner}},
  \bibinfo {author} {\bibfnamefont {A.}~\bibnamefont {Stolz}}, \ and\ \bibinfo
  {author} {\bibfnamefont {I.}~\bibnamefont {Wiedenhoever}},\ }\href@noop {}
  {\bibfield  {journal} {\bibinfo  {journal} {Nucl. Instrum. Meth. A}\ }\textbf
  {\bibinfo {volume} {204}},\ \bibinfo {pages} {90} (\bibinfo {year}
  {2003})}\BibitemShut {NoStop}%
\bibitem [{\citenamefont {Bird}\ \emph {et~al.}(2005)\citenamefont {Bird},
  \citenamefont {Kenney}, \citenamefont {Toth}, \citenamefont {Weijers},
  \citenamefont {DeKamp}, \citenamefont {Thoennessen},\ and\ \citenamefont
  {Zeller}}]{SWEEPER}%
  \BibitemOpen
  \bibfield  {author} {\bibinfo {author} {\bibfnamefont {M.~D.}\ \bibnamefont
  {Bird}}, \bibinfo {author} {\bibfnamefont {S.~J.}\ \bibnamefont {Kenney}},
  \bibinfo {author} {\bibfnamefont {J.}~\bibnamefont {Toth}}, \bibinfo {author}
  {\bibfnamefont {H.~W.}\ \bibnamefont {Weijers}}, \bibinfo {author}
  {\bibfnamefont {J.~C.}\ \bibnamefont {DeKamp}}, \bibinfo {author}
  {\bibfnamefont {M.}~\bibnamefont {Thoennessen}}, \ and\ \bibinfo {author}
  {\bibfnamefont {A.~F.}\ \bibnamefont {Zeller}},\ }\href@noop {} {\bibfield
  {journal} {\bibinfo  {journal} {IEEE Trans. Appl. Supercond.}\ }\textbf
  {\bibinfo {volume} {15}},\ \bibinfo {pages} {1252} (\bibinfo {year}
  {2005})}\BibitemShut {NoStop}%
\bibitem [{\citenamefont {Christian}\ \emph
  {et~al.}(2012{\natexlab{a}})\citenamefont {Christian} \emph
  {et~al.}}]{Chr12}%
  \BibitemOpen
  \bibfield  {author} {\bibinfo {author} {\bibfnamefont {G.}~\bibnamefont
  {Christian}} \emph {et~al.},\ }\href@noop {} {\bibfield  {journal} {\bibinfo
  {journal} {Phys. Rev. C}\ }\textbf {\bibinfo {volume} {85}},\ \bibinfo
  {pages} {034327} (\bibinfo {year} {2012}{\natexlab{a}})}\BibitemShut
  {NoStop}%
\bibitem [{\citenamefont {Kohley}\ \emph {et~al.}(2012)\citenamefont {Kohley},
  \citenamefont {Lunderberg}, \citenamefont {DeYoung}, \citenamefont {Roeder},
  \citenamefont {Baumann}, \citenamefont {Christian}, \citenamefont {Mosby},
  \citenamefont {Smith}, \citenamefont {Snyder}, \citenamefont {Spyrou},\ and\
  \citenamefont {Thoennessen}}]{Koh12}%
  \BibitemOpen
  \bibfield  {author} {\bibinfo {author} {\bibfnamefont {Z.}~\bibnamefont
  {Kohley}}, \bibinfo {author} {\bibfnamefont {E.}~\bibnamefont {Lunderberg}},
  \bibinfo {author} {\bibfnamefont {P.~A.}\ \bibnamefont {DeYoung}}, \bibinfo
  {author} {\bibfnamefont {B.~T.}\ \bibnamefont {Roeder}}, \bibinfo {author}
  {\bibfnamefont {T.}~\bibnamefont {Baumann}}, \bibinfo {author} {\bibfnamefont
  {G.}~\bibnamefont {Christian}}, \bibinfo {author} {\bibfnamefont
  {S.}~\bibnamefont {Mosby}}, \bibinfo {author} {\bibfnamefont {J.~K.}\
  \bibnamefont {Smith}}, \bibinfo {author} {\bibfnamefont {J.}~\bibnamefont
  {Snyder}}, \bibinfo {author} {\bibfnamefont {A.}~\bibnamefont {Spyrou}}, \
  and\ \bibinfo {author} {\bibfnamefont {M.}~\bibnamefont {Thoennessen}},\
  }\href@noop {} {\bibfield  {journal} {\bibinfo  {journal} {Nucl. Instrum.
  Meth. Phys. Res. A}\ }\textbf {\bibinfo {volume} {682}},\ \bibinfo {pages}
  {59} (\bibinfo {year} {2012})}\BibitemShut {NoStop}%
\bibitem [{\citenamefont {Volya}\ and\ \citenamefont
  {Zelevinsky}(2006)}]{Volya06}%
  \BibitemOpen
  \bibfield  {author} {\bibinfo {author} {\bibfnamefont {A.}~\bibnamefont
  {Volya}}\ and\ \bibinfo {author} {\bibfnamefont {V.}~\bibnamefont
  {Zelevinsky}},\ }\href@noop {} {\bibfield  {journal} {\bibinfo  {journal}
  {Phys. Rev. C}\ }\textbf {\bibinfo {volume} {74}},\ \bibinfo {pages} {064314}
  (\bibinfo {year} {2006})}\BibitemShut {NoStop}%
\bibitem [{\citenamefont {Schmidt}\ \emph {et~al.}(1993)\citenamefont
  {Schmidt}, \citenamefont {Morrislon},\ and\ \citenamefont
  {Witherell}}]{Sch93}%
  \BibitemOpen
  \bibfield  {author} {\bibinfo {author} {\bibfnamefont {D.}~\bibnamefont
  {Schmidt}}, \bibinfo {author} {\bibfnamefont {R.}~\bibnamefont {Morrislon}},
  \ and\ \bibinfo {author} {\bibfnamefont {M.}~\bibnamefont {Witherell}},\
  }\href@noop {} {\bibfield  {journal} {\bibinfo  {journal} {Nucl. Instrum.
  Meth. A}\ }\textbf {\bibinfo {volume} {328}},\ \bibinfo {pages} {547}
  (\bibinfo {year} {1993})}\BibitemShut {NoStop}%
\bibitem [{\citenamefont {Christian}\ \emph
  {et~al.}(2012{\natexlab{b}})\citenamefont {Christian} \emph
  {et~al.}}]{Chr12PRL}%
  \BibitemOpen
  \bibfield  {author} {\bibinfo {author} {\bibfnamefont {G.}~\bibnamefont
  {Christian}} \emph {et~al.},\ }\href@noop {} {\bibfield  {journal} {\bibinfo
  {journal} {Phys. Rev. Lett.}\ }\textbf {\bibinfo {volume} {108}},\ \bibinfo
  {pages} {032501} (\bibinfo {year} {2012}{\natexlab{b}})}\BibitemShut
  {NoStop}%
\end{thebibliography}

\end{document}